\documentclass[useAMS,usenatbib]{mnras}
\usepackage{times}
\usepackage{graphicx}

\usepackage{amsmath}
\usepackage{amssymb}

\usepackage[T1]{fontenc}
\usepackage{aecompl}
\usepackage{epstopdf}

\begin{document}

\title[Massive Pop III Galaxies]{What is the Maximum Mass of a Population III Galaxy?}

\author[E. Visbal et al.]{Eli Visbal$^{1,2}$\thanks{evisbal@flatironinstitute.org}, Greg L. Bryan$^{1,2}$, Zolt\'{a}n Haiman$^{2, 3}$ \\ $^1$Center for Computational Astrophysics, Flatiron Institute, 162 5th Ave, New York, NY, 10003, U.S.A. \\ 
$^2$Department of Astronomy, Columbia University, 550 West 120th Street, New York, NY, 10027, U.S.A. \\
$^3$Center for Cosmology and Particle Physics, New York University, 4 Washington Place, New York, NY, 10003, U.S.A. }

\maketitle

\begin{abstract}
We utilize cosmological hydrodynamic simulations to study the formation of Population III (Pop III) stars in dark matter halos exposed to strong ionizing radiation. We simulate the formation of three halos subjected to a wide range of ionizing fluxes, and find that for high flux, ionization and photoheating can delay gas collapse and star formation up to halo masses significantly larger than the atomic cooling threshold. The threshold halo mass at which gas first collapses and cools increases with ionizing flux for intermediate values, and saturates at a value approximately an order of magnitude above the atomic cooling threshold for extremely high flux (e.g. $\approx 5 \times 10^8 ~ M_\odot$ at $z\approx6$). This behavior can be understood in terms of photoheating, ionization/recombination, and Ly$\alpha$ cooling in the pressure-supported, self-shielded gas core at the center of the growing dark matter halo. We examine the spherically-averaged radial velocity profiles of collapsing gas and find that a gas mass of up to $\approx 10^{6}~ M_\odot$  can reach the central regions within $3~{\rm Myr}$, providing an upper limit on the amount of massive Pop III stars that can form. The ionizing radiation increases this limit by a factor of a few compared to strong Lyman-Werner (LW) radiation alone. We conclude that the bright HeII 1640 \AA\ emission recently observed from the high-redshift galaxy CR7 cannot be explained by Pop III stars alone. However, in some halos, a sufficient number of Pop III stars may form to be detectable with future telescopes such as the \emph{James Webb Space Telescope} (JWST).
\end{abstract}


\section{Introduction}
Cosmological simulations predict that the first metal-free Pop III stars form in $\sim 10^{5-6} M_{\odot}$
dark matter ``minihalos'' in the early Universe \citep[for a recent review, see][]{2015ComAC...2....3G}. Molecular hydrogen (${\rm H}_2$) is essential for star formation in these halos because their virialized gas does not reach sufficient temperatures to cool through atomic hydrogen transitions.  Hydrogen molecules  can be photodissociated by LW radiation (11.2-13.6 eV photons), and as the star formation density of the Universe increases,
a LW background builds up which suppresses star formation in small minihalos \citep[e.g.][]{1997ApJ...476..458H,2001ApJ...548..509M,2007ApJ...671.1559W,2008ApJ...673...14O,2011MNRAS.418..838W,2014MNRAS.445..107V}. Eventually, in regions with sufficiently strong LW radiation, Pop III star formation is suppressed in all halos up to the virial temperature where atomic cooling becomes efficient, $T_{\rm vir} \approx 10^4 ~{\rm K}$, corresponding to a redshift-dependent halo mass of $M \approx 3 \times 10^7 \left ( \frac{1+z}{11}\right )^{-3/2} M_\odot$ \citep{2001PhR...349..125B}. Preventing star formation up to this mass with LW radiation alone requires a very high flux (Regan et al. 2016, submitted). LW feedback along with inefficient metal mixing may result in significant Pop III star formation during the epoch of reionization or even at much later times \citep{2006Natur.440..501J, 2007ApJ...659..890W, 2009ApJ...694..879T}.

Generally it has been thought that Pop III stars do not form in halos much larger than the atomic cooling threshold.
This is because the lifetimes of massive ($M \gtrsim 100 ~ M_\odot$) Pop III stars are only a few Myr \citep{2002A&A...382...28S}, leading to prompt enrichment of metals via supernovae winds and a transition to Population II (Pop II) stars. However, it has been proposed that Pop III stars could form in much more massive halos if star formation is suppressed in all progenitor halos via strong ionizing radiation \citep{2010MNRAS.404.1425J, 2016MNRAS.460L..59V, 2016arXiv161004249Y}.

This was suggested recently as a possible interpretation of the high-redshift Ly$\alpha$ emitter ``CR7'' \citep{2015ApJ...808..139S, 2015MNRAS.453.2465P, 2016MNRAS.460L..59V}, which has a strong HeII 1640~\AA ~recombination line that could be produced by massive Pop III stars. Using the stellar models of \cite{2002A&A...382...28S} (assuming an ionizing escape fraction of zero, gas temperature of 30,000 K, and electron density of $n_e = 100~{\rm cm^{-3}}$), we find that total stellar masses of $\sim 3\times 10^7~M_\odot$ or $\sim 10^7 M_\odot$ are required to generate the HeII luminosity of CR7 \citep{2015ApJ...808..139S} if all stars are $120~M_\odot$ or $1000~M_\odot$ zero age main sequence Pop III stars, respectively. In addition to correctly interpreting CR7, understanding this mode of Pop III star formation is important because it may lead to a larger, and thus more easily observable, total stellar mass in Pop III galaxies. Massive Pop III-forming halos would be promising targets for next-generation telescopes, such as \emph{JWST}, leading to the first detection of metal-free stars. 

In this paper, we utilize cosmological hydrodynamics simulations to investigate photoionization feedback on pristine dark matter halos, leading to Pop III star formation in halos significantly above the atomic cooling threshold. Previous work has investigated the photoevaporation of minihalos due to ionizing radiation \citep{2005MNRAS.361..405I,2004MNRAS.348..753S}. There has also been a large body of work seeking to understand the suppression of star formation in more massive dark matter halos due to reionization \citep[e.g.][]{1994ApJ...427...25S,1996ApJ...465..608T,1998MNRAS.296...44G,2000ApJ...542..535G,2004ApJ...601..666D,2006MNRAS.371..401H,2008MNRAS.390..920O,2013MNRAS.432L..51S,2014MNRAS.444..503N}. \cite{2004ApJ...601..666D} find, using 1D, spherically symmetric simulations, that photoionization feedback is not important in halos with circular velocity $v_{\rm c}\gtrsim 10~{\rm km~s^{-1}}$, when subjected to the mean ultraviolet background at $z\gtrsim 10$. Here we consider much higher ionizing fluxes, as might be expected from nearby star-forming galaxies, which results in feedback being important at significantly larger halo masses. 

Much of the previous work assumes that the key physical scale in the problem is the Jeans mass of ionized photoheated gas at the mean cosmic density, or a related, time-averaged version of this quantity  \citep[the so-called filtering mass;][]{1998MNRAS.296...44G}. Recently, \cite{2014MNRAS.444..503N} have put forward a physical picture that goes beyond this approximation, considering the effects of gravity, cooling, pressure, and self-shielding on gas as it collapses into a halo. This model is able to reproduce the gas mass fraction within dark matter halos as a function of halo mass and redshift. We note that, in general, the goal of these previous works was to find either the characteristic halo mass at which gas accretion can occur or the mass at which halos contain a gas fraction of 50 per cent (relative to the total cosmic mean ratio of baryons to dark matter). In this paper, we pose and address a closely related, yet distinct question: {\em At what halo mass can \emph{any} gas first cool and form stars in a pristine halo subjected to ionizing radiation?}

Naively, one might expect that if gas is ionized before any star formation occurs, pressure will prevent it from collapsing and forming stars before it reaches the Jeans/filtering mass at the mean gas density of the Universe. We find that modeling the delay of star formation in pristine dark matter halos considering only the Jeans/filtering scale does not provide a complete physical picture. Instead, the essential physics can be described as follows. After gas is photoionized and photoheated, it forms a pressure supported core inside the center of the forming dark matter halo. As the halo grows, the density of this core increases and recombinations increase the neutral fraction. Eventually, the heating timescale in the self-shielded core becomes significantly longer than the dynamical timescale, leading to runaway collapse and star formation. The halo mass at which this occurs depends on the intensity of the ionizing flux. For extremely high flux, we find masses roughly an order of magnitude higher than the atomic cooling threshold.

The simulations presented below quantify the mass at which star formation first occurs, for several halos across a wide range of photoionizing fluxes. For the three halos tested, we find that an ionizing photon flux of at least $\approx 10^6 ~{\rm s^{-1} ~ cm^{-2}}$ is required to suppress star formation significantly above the atomic cooling threshold. At high-redshift, this is most likely to occur in close proximity to large star-forming galaxies \citep{2010MNRAS.404.1425J, 2016MNRAS.460L..59V}.

We examine the radial velocity profiles of gas in our simulations to investigate the total mass of Pop III stars formed. We find an upper limit of $\approx 10^6~M_\odot$ within $3~{\rm Myr}$ after collapse (corresponding to the lifetime of massive Pop III stars). This limit varies by a factor of a few depending on the halo and applied ionizing flux (it may also be decreased by a factor of a few by molecular hydrogen cooling, as discussed below). Even though the mass of halos hosting Pop III stars can be increased by more than an order of magnitude, the increase in limits on stellar mass is found to be more modest (a factor of a few increase for very high flux). As discussed below, we find that these limits suggest that CR7 cannot be explained solely by Pop III stars. Nevertheless, the maximum total stellar masses of Pop III galaxies we infer are large enough to be observable with \emph{JWST}.

This paper is structured as follows. In \S2, we discuss the details of our cosmological simulations. We explain the physics of delayed star formation in a pristine halo subjected to strong ionizing radiation in \S3. In \S4, we use our simulations to approximate an upper limit on the amount of gas which could form Pop III stars once a halo is massive enough to overcome photoionization feedback. Finally, we summarize our results and conclusions in \S5. Throughout we assume a $\Lambda$CDM cosmology consistent with \cite{2014A&A...571A..16P}: $\Omega_{\rm m} = 0.32$, $\Omega_{\Lambda} = 0.68$, $\Omega_{\rm b} = 0.049$, $h=0.67$, $\sigma_8=0.83$, and $n_{\rm s} = 0.96$.

\section{Simulations}
\subsection{Basic Setup}
We employ cosmological hydrodynamics simulations performed with the adaptive mesh refinement code \textsc{enzo} \citep{2014ApJS..211...19B}, to study the delay of Pop III star formation in dark matter halos subjected to strong ionizing radiation. We simulate three different halos (``halos A, B, and C'') which were first identified in dark matter only simulations and then re-simulated with ``zoom-in'' runs where only the region of halo formation is refined. The initial conditions for all of the simulations were computed with the \textsc{music} software package \citep{2011MNRAS.415.2101H}. Our zoom-in initial conditions each contain three nested grids.

Halo A is taken from an $L=3.5 h^{-1}~ {\rm Mpc}$ box and the most refined region is 0.13 times this length. Halos B and C are taken from an $L=2 h^{-1}~ {\rm Mpc}$ box and the lengths of the most refined regions are 0.18 and 0.16 times the total box length, respectively.  Halos A, B, and C span a range of masses at fixed redshift. When no ionizing flux is included, they reach the atomic cooling threshold at $z\approx 21$, $15$, and $12$ respectively. For all of our runs, the full box has cell and particle resolutions of $128^3$. 
For most of our runs, the most refined region has an initial resolution of $512^3$ and a maximum refinement level of 18 for gas and 13 for dark matter particles. 
This resolution corresponds to a dark matter particle mass of $36000$, $7000$, and $7000~ M_\odot$ and an initial baryon cell mass of $6500$, $1200$, and $1200~ M_\odot$ for halos A, B, and C, respectively. For a few high-resolution convergence checks, we increase the initial mass resolution of the refined region by a factor of 8.

\subsection{Treatment of Ionizing Radiation}
We treat ionizing radiation in our simulations with an approximate method that we have incorporated into \textsc{enzo} version 2.5. This treatment is based on \cite{2013MNRAS.430.2427R}, which provides fitting functions to the photoionization rate as a function of local density that match cosmological simulations with radiative transfer in the post-reionization Universe. We assume a uniform photoionization rate throughout the entire box, modified by a local self-shielding factor estimated from the temperature and total (neutral+ionized) hydrogen number density.

The values of the photoionization rate are then given by the following relation:
\begin{equation} 
\Gamma = f_{\rm sh} \Gamma_{\rm bg},
\end{equation}
where $\Gamma_{\rm bg}$ is the photoionization rate due to a uniform ionizing background without shielding and $f_{\rm sh}$ is the local shielding factor which depends on the temperature and hydrogen density,
\begin{equation}
\label{shield_eqn}
f_{\rm sh} = 0.98 \left [ 1 + \left (   \frac{n_{\rm H}}{n_{\rm H, sh}} \right )^{1.64} \right ]^{-2.28}  + 0.02 \left [ 1 + \frac{n_{\rm H}}{n_{\rm H, sh}} \right ]^{-0.84},
\end{equation}
where
\begin{equation}
n_{\rm H, sh} = 5 \times 10^{-3} {\rm cm}^{-3} \left ( \frac{T}{10^4 ~{\rm K}} \right )^{0.17} \left ( \frac{\Gamma_{\rm bg}}{10^{-12} ~{\rm s}^{-1}} \right )^{2/3}.
\end{equation}
The same shielding factor is applied to photoheating as a result of hydrogen ionization. When computing the relative hydrogen ionization and heating rates, we assume a
$T_*=3\times 10^4 ~{\rm K}$ black-body spectrum. We do not include any helium photoionization or photoheating. For the spectra of Pop II stars, we do not expect that including helium would change our results. However, for harder spectra (e.g. from active galactic nuclei), double ionization of helium could potentially heat ionized gas to higher temperatures, leading to larger collapse masses than discussed below. We note that the functional form of the shielding factor is calibrated in \cite{2013MNRAS.430.2427R} with cosmological simulations including radiative transfer at lower redshifts than we explore here. However, as we demonstrate next, this prescription gives similar results to using the (computationally much more expensive) ray-tracing radiative transfer available in \textsc{enzo}.

The main goal of this paper is to determine how the halo mass and redshift when gas first collapses and forms stars depends on the intensity of ionization radiation. To test the accuracy of our approximate method, we compare a simulation of halo A, performed with the method described above, to a simulation which includes radiative transfer. For the simulation with radiative transfer, we utilize \textsc{enzo} version 3.0 with the ray tracing radiative transfer algorithm described in \citep{2011MNRAS.414.3458W}. 
When performing the simulation with radiative transfer, we insert a radiation particle at the center of the box at $z=30$ which emits $7.5 \times 10^{54}$ photons per second.  The photon energies are in 6 bins spaced evenly in log-space such that 10 per cent of these photons are above the hydrogen ionization threshold. This setup is meant to resemble a halo forming in pristine gas exposed to ionizing radiation from a large Pop II galaxy/galaxies in a large HII bubble. In the case of the local shielding approximation, we simulate the same halo with a uniform ionizing background of $1.3 \times 10^8 {\rm photons~s^{-1} cm^{-2}}$ (also turned on at $z=30$). The background is chosen to roughly match the ionizing flux at the location of halo collapse (at $z\approx 14$), in the simulation with radiative transfer. This collapse location is approximately $7~{\rm  kpc}$ from the location of the radiation particle at the time of collapse. For this comparison example, we have chosen a large flux to test our approximate method in the high-flux case. We do not repeat the same comparison for other values of the flux, due to the high computational cost of utilizing radiation transfer.  

\begin{figure}
\includegraphics[width=90mm]{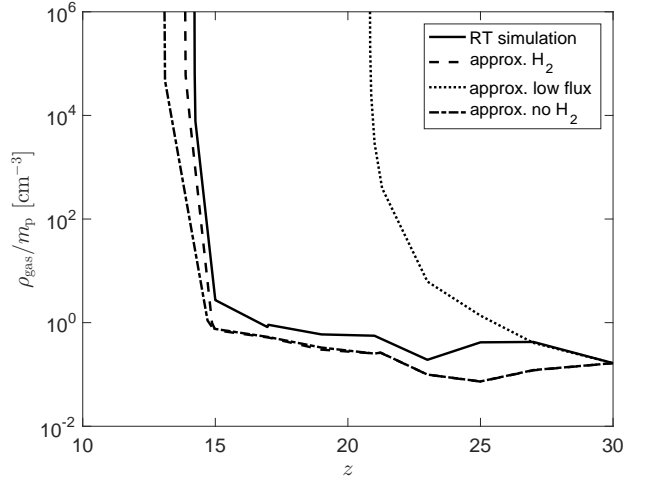}
\caption{\label{RT_fig} The maximum gas density in halo A as a function of redshift. The three curves showing runaway collapse at $z\approx 14$ are for a strong ionizing source, with ionizing flux of $\sim  10^8 ~{\rm photons~s^{-1} cm^{-2}}$. The solid curve is simulated with \textsc{enzo}'s ray tracing treatment of radiative transfer, while the dashed and dot-dashed curves are computed with the simpler self-shielding approximation discussed in \S2.2. This demonstrates that the simpler and less numerically expensive prescription gives a consistent collapse redshift with the radiative transfer run. The dotted curve is for flux 2000 times lower, which results in collapse at the atomic cooling threshold much earlier on. The dashed and dot-dashed curves include and exclude molecular hydrogen cooling, respectively (as discussed in \S2.3). }
\end{figure}

Figure \ref{RT_fig} shows the maximum gas density within Halo A as a function of redshift for both simulation methods. We emphasize that the radiative transfer simulation gives very similar results to the much less computationally demanding shielding approximation. For the remainder of this work, we exclusively discuss runs performed with this approximate method.
 For comparison we also plot the shielding approximation case with 2000 times lower flux, leading to collapse at $T_{\rm vir} \approx 10^4~{\rm K}$. The higher ionizing flux significantly delays gas collapse and star formation.

\subsection{Molecular Hydrogen}
\textsc{enzo} has options to include a chemical network with 6 species (${\rm H}$, ${\rm He}$, ${\rm H^+}$, ${\rm He^+}$, ${\rm He^{++}}$, and ${\rm e}^-$), as well as an expanded network which also includes ${\rm H_2}$, ${\rm H^-}$, and ${\rm H_2^+}$. The main focus of this paper is to understand the impact of ionizing radiation on Pop III star formation assuming that LW radiation can suppress star formation up to the atomic cooling threshold. For this reason, we do not perform a detailed study of the effect of molecules, however we briefly discuss their impact here. 

For the comparison shown in Figure \ref{RT_fig}, the case with radiative transfer includes molecular hydrogen as well as a uniform LW background with a value $J_{\rm LW} = 100 \times J_{21}$, defined such that the ${\rm H_2}$ photodissociation rate equals $ k_{\rm H_2}  =  1.42 \times 10^{-12}  J_{\rm LW} / J_{21} ~ {\rm s^{-1}} $ and $J_{21} = 10^{-21}~ {\rm erg~s^{-1}~cm^{-2}~Hz^{-1}~sr^{-1}}$. This photodissociation rate is modified by the self-shielding function described in \cite{2011MNRAS.418..838W}. We run the simulation both with molecular hydrogen and the same LW background as well as a case without any molecular hydrogen. We find that including LW feedback and molecules results in approximately the same collapse redshift as not including any molecules. 

When performing similar simulations on halos B and C with the shielding approximation and molecular hydrogen, we find that turning on ionization at $z=20$ actually enhances ${\rm H_2}$ formation and leads to collapse before the halo reaches the atomic cooling threshold (even with a $J_{\rm LW}=100 \times J_{21}$ background). We suspect that 
our assumption of a spatially uniform and isotropic photoionizing background may be artificially promoting collapse, since the outer regions ionize and heat first, simultaneously compressing the halo from all sides unrealistically. This does not occur in our examples with halo A above because the ionizing radiation is turned on before a minihalo has formed. Ignoring this likely unrealistic effect, collapse should be triggered by H line cooling. For this reason, for the rest of our simulations we do not include any molecular hydrogen. The results in Figure \ref{RT_fig} suggest that this should have a relatively small impact on the redshift and halo mass of collapse. However, molecular hydrogen may have a larger impact on our estimated limits of Pop III stellar mass formed, as we discuss in \S4.

\section{Ionization Feedback}
\begin{figure*}
\includegraphics[width=80mm]{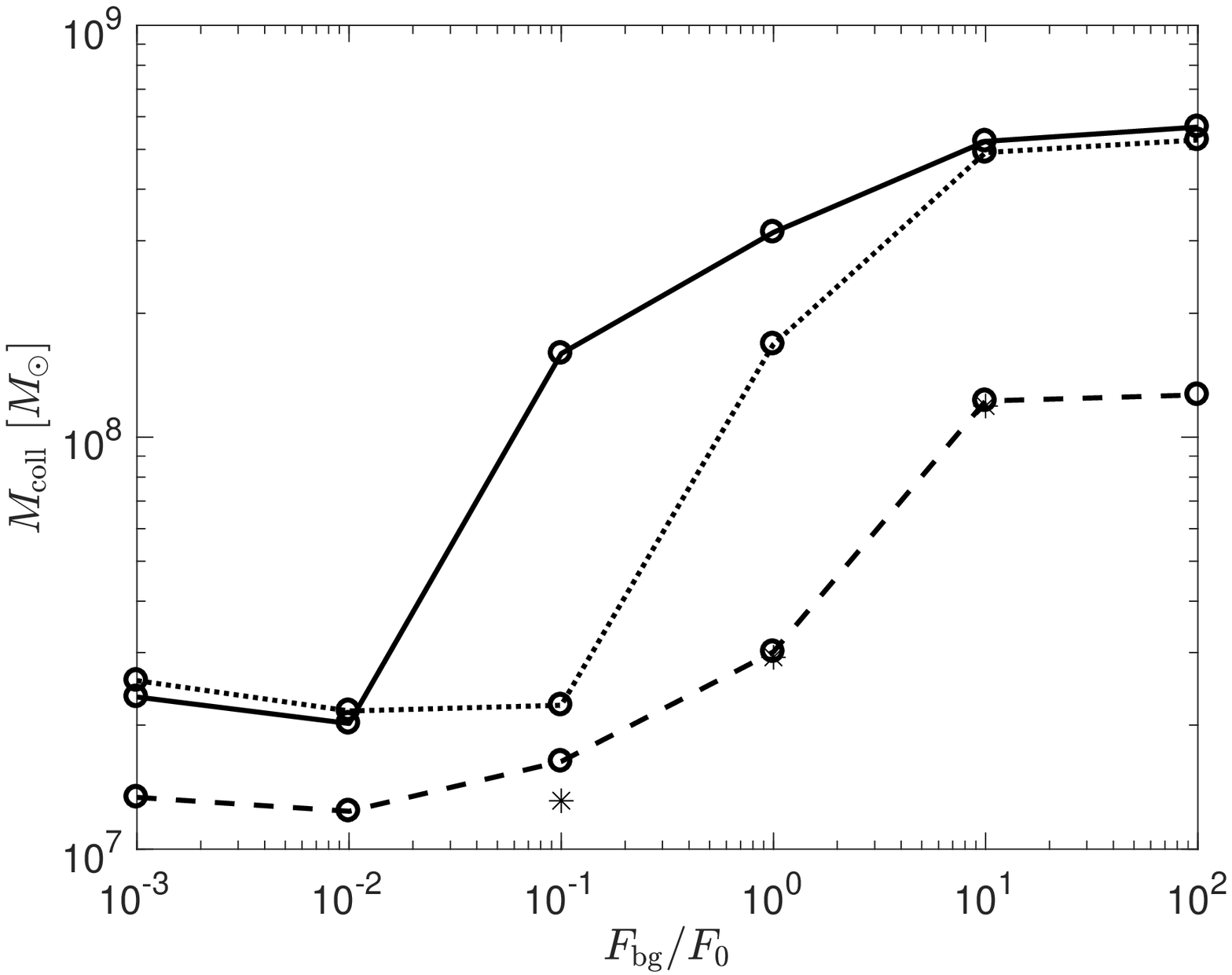}
\includegraphics[width=80mm]{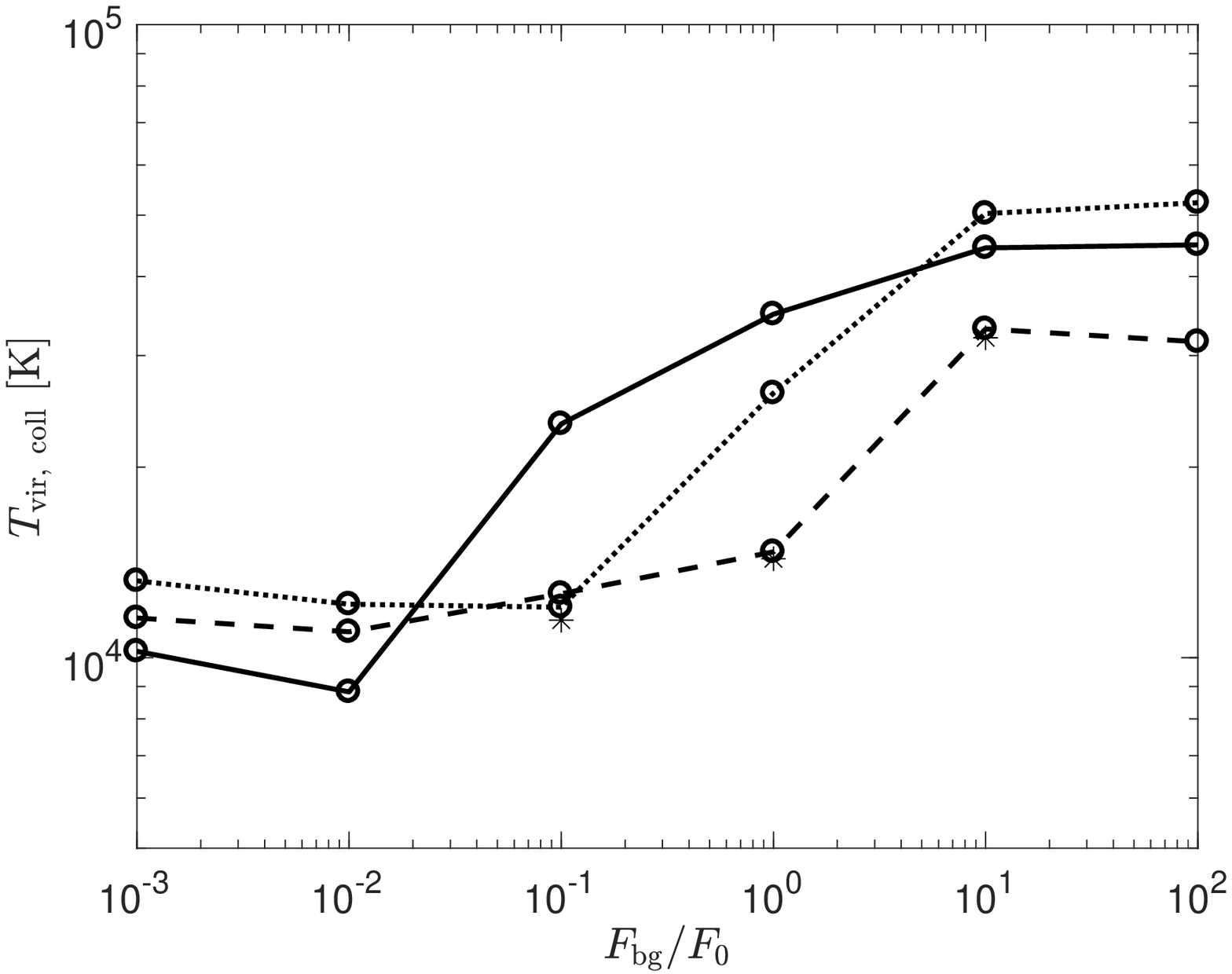}
\includegraphics[width=80mm]{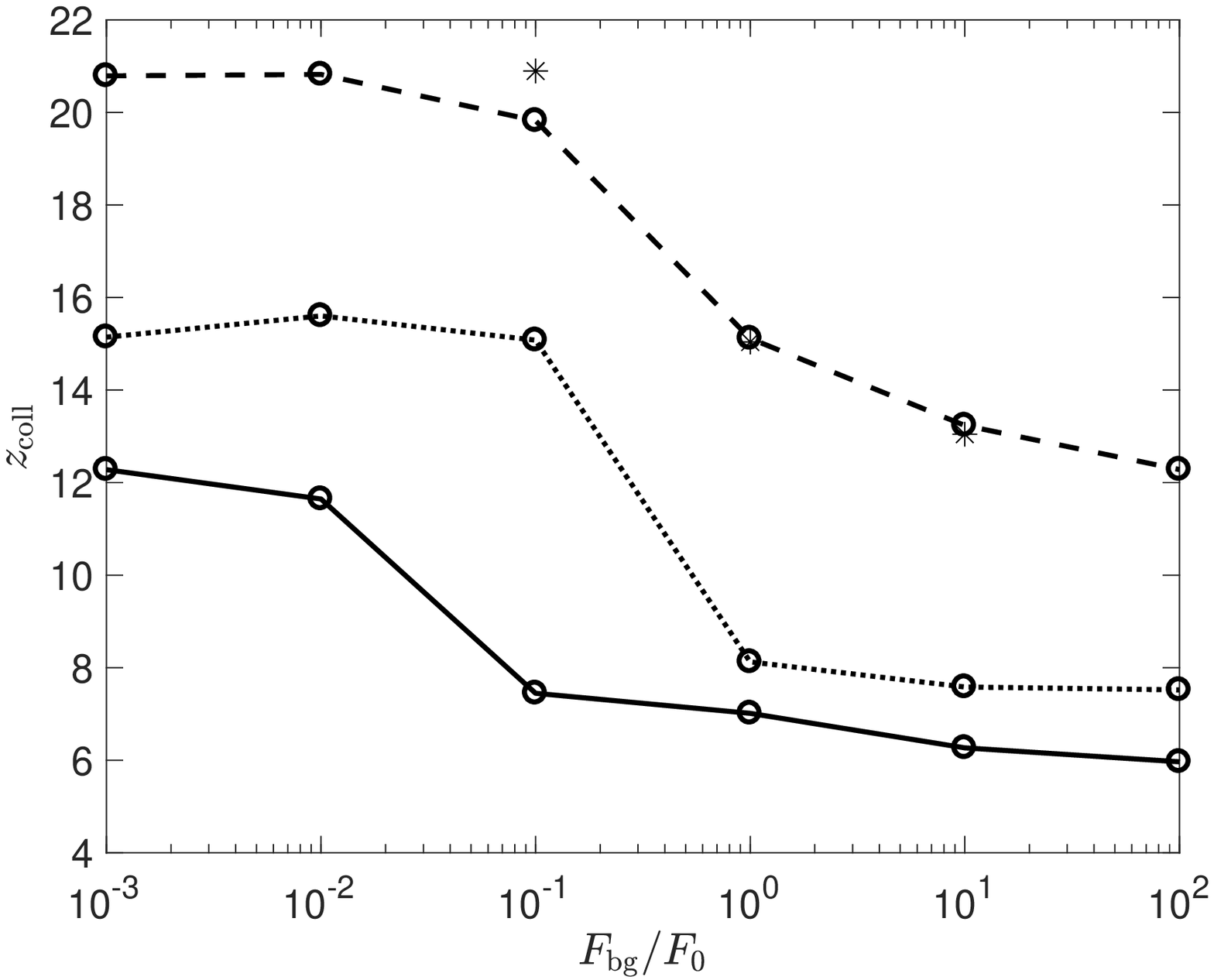}
\caption{\label{m_fig} The halo mass, redshift, and virial temperature when runaway gas cooling and collapse first occur in each halo, shown as a function of ionizing flux. The dashed, dotted, and solid lines represent halos A, B, and C respectively (with open circles denoting the actual simulation data). The stars (*) are for halo A, but with 8 times higher resolution in the refinement region.  }
\end{figure*}

In order to characterize the impact of photoionization feedback on Pop III star formation, we simulate the formation of halos A, B, and C over a wide range of ionizing fluxes. We run each simulation until it reaches the maximum level of refinement during the runaway cooling and collapse of gas, which should immediately precede star formation. In Figure \ref{m_fig}, we plot the mass, redshift, and virial temperature at collapse for each halo as a function of ionizing flux. 
For simplicity, we plot the virial temperature as defined in \cite{2001PhR...349..125B} assuming $\mu=1.22$, which corresponds to neutral gas (even though for high fluxes the gas is ionized).
We quote all flux values in terms of a reference flux corresponding to $2\times 10^{54}$ ionizing photons per second from a source at a distance of $50~{\rm kpc}$,  $F_0 = 6.7\times 10^6~ {\rm photons~s^{-1}~cm^{-2}}$. We note the merger tree analysis from \cite{2016MNRAS.460L..59V} demonstrates that a $M = 6.6 \times 10^{11} ~ M_\odot$ dark matter halo at $z \approx 7$ with a 10 per cent star formation efficiency and an ionizing photon escape fraction of $f_{\rm esc}=0.1$ will produce $\sim 2 \times 10^{53}$ ionizing photons per second over a range of $\Delta z \approx 10 $ in redshift (see their Figure 1). Thus, at a separation of $\sim 50~ {\rm kpc}$ from such a halo, the flux will be $0.1F_0$ up to very high redshift. 

For halo A, the flux is turned on at $z=30$ and for halos B and C the flux is turned on at $z=20$. We find that as long as there is sufficient time for the intergalactic medium (IGM) to be reionized before halo formation, changing the start time of the source does not have a large impact on our results. For example, we performed a run with $F_0$ flux starting at $z=16$ on halo C and found collapse at almost exactly the same redshift as the case starting at $z=20$. To check for convergence, we simulated halo B with mass resolution 8 times higher in the refinement region than our other runs. In Figure \ref{m_fig}, we plot both the low- and high-resolution results for collapse time, redshift and virial temperature. The close match between the different resolutions suggests that our results have converged.

For the lowest fluxes, collapse happens at $T_{\rm vir}\approx 10^4~{\rm K}$ for all of our halos. At intermediate flux (the precise intensity varies for each halo), the collapse mass increases with flux, and for high flux the mass dependence on flux again becomes very weak. Examining our runs in detail gives us a clear picture of the physical processes resulting in this behavior. 

The low flux collapse at  $T_{\rm vir}\approx 10^4~{\rm K}$ is simply because this is the lowest temperature at which atomic hydrogen cooling is efficient (as discussed above, we assume molecules have been dissociated by LW radiation). Gas in smaller halos cannot cool and therefore does not undergo runaway collapse. 

\begin{figure*}
\includegraphics[width=80mm]{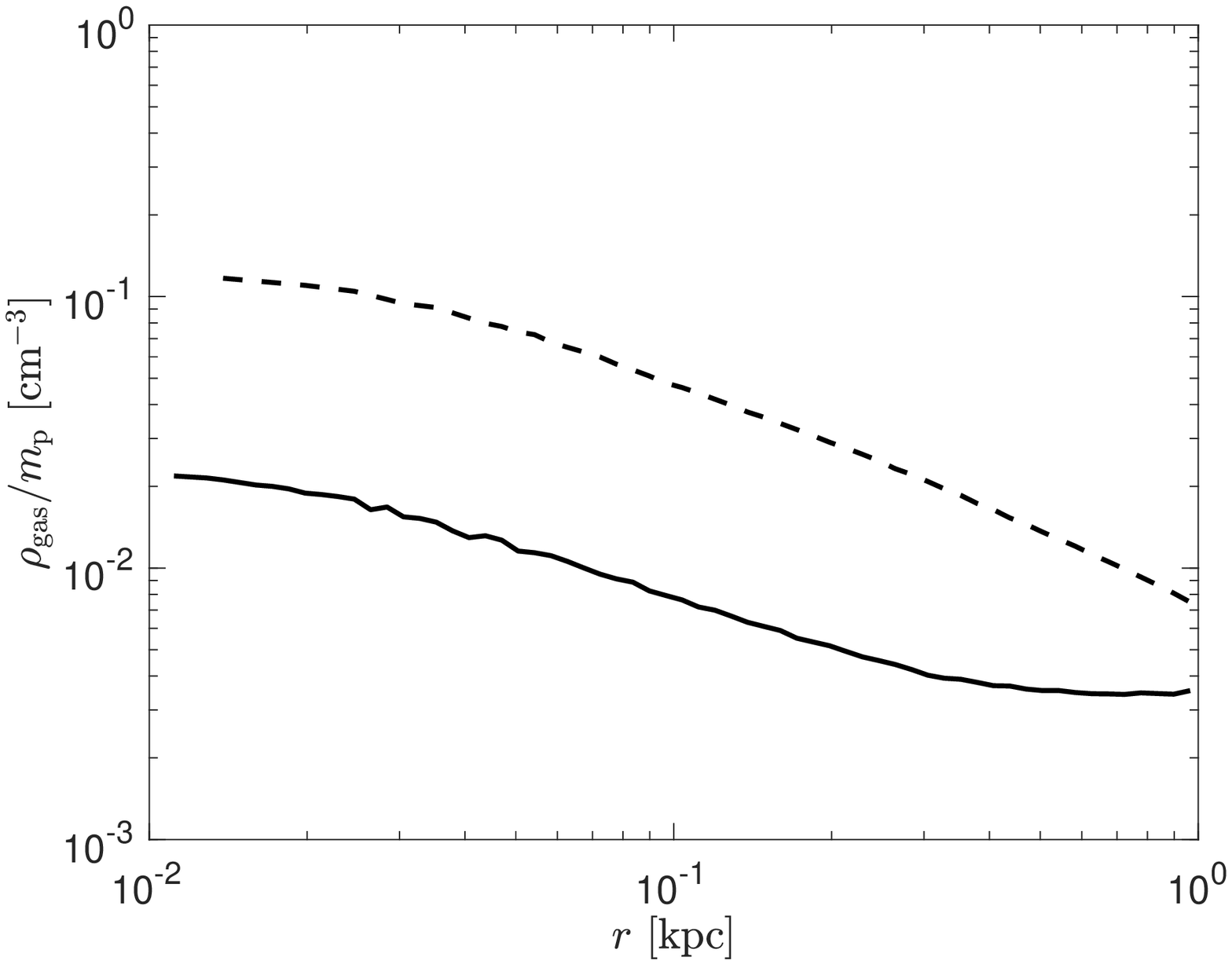}
\includegraphics[width=80mm]{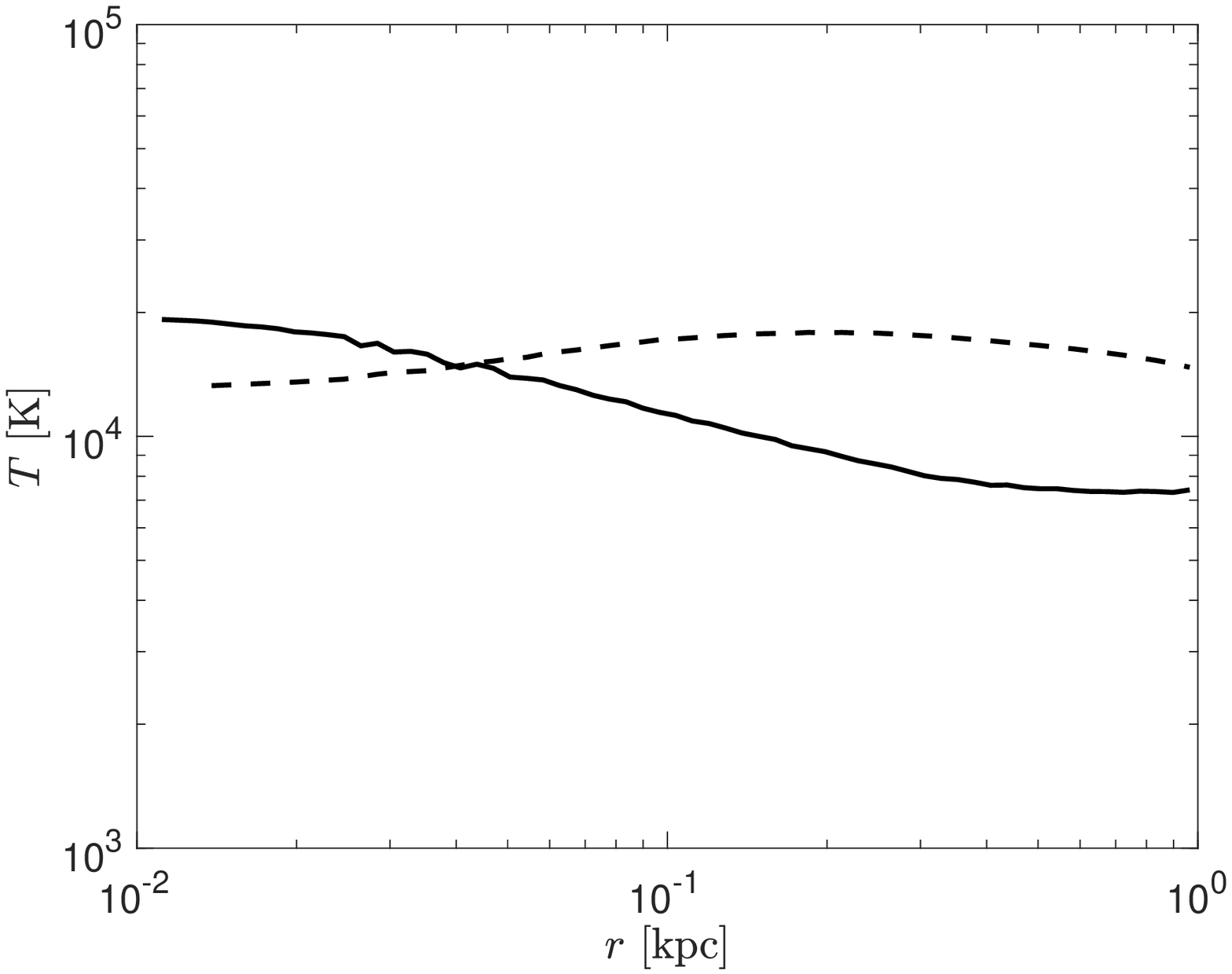}
\includegraphics[width=80mm]{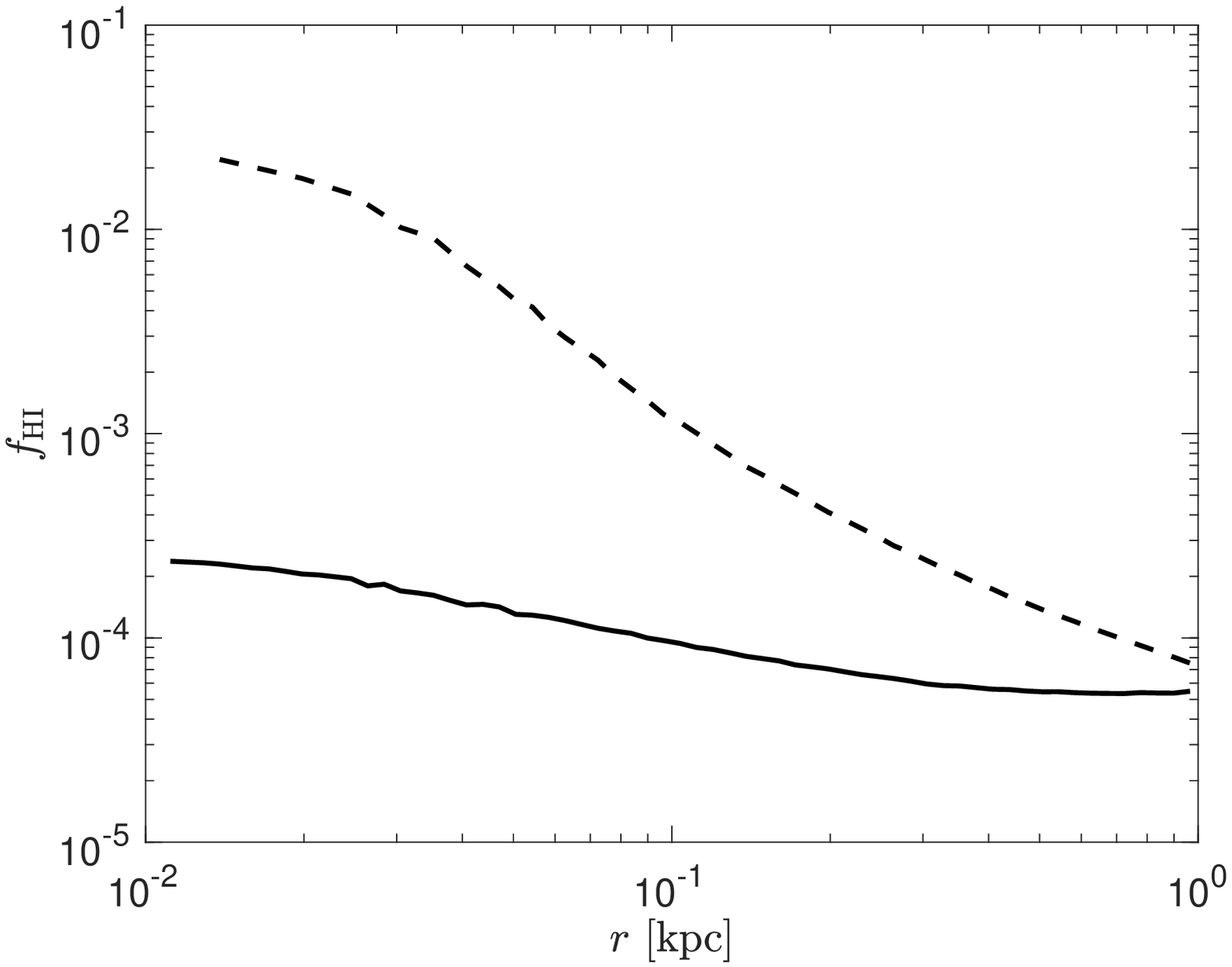}
\includegraphics[width=80mm]{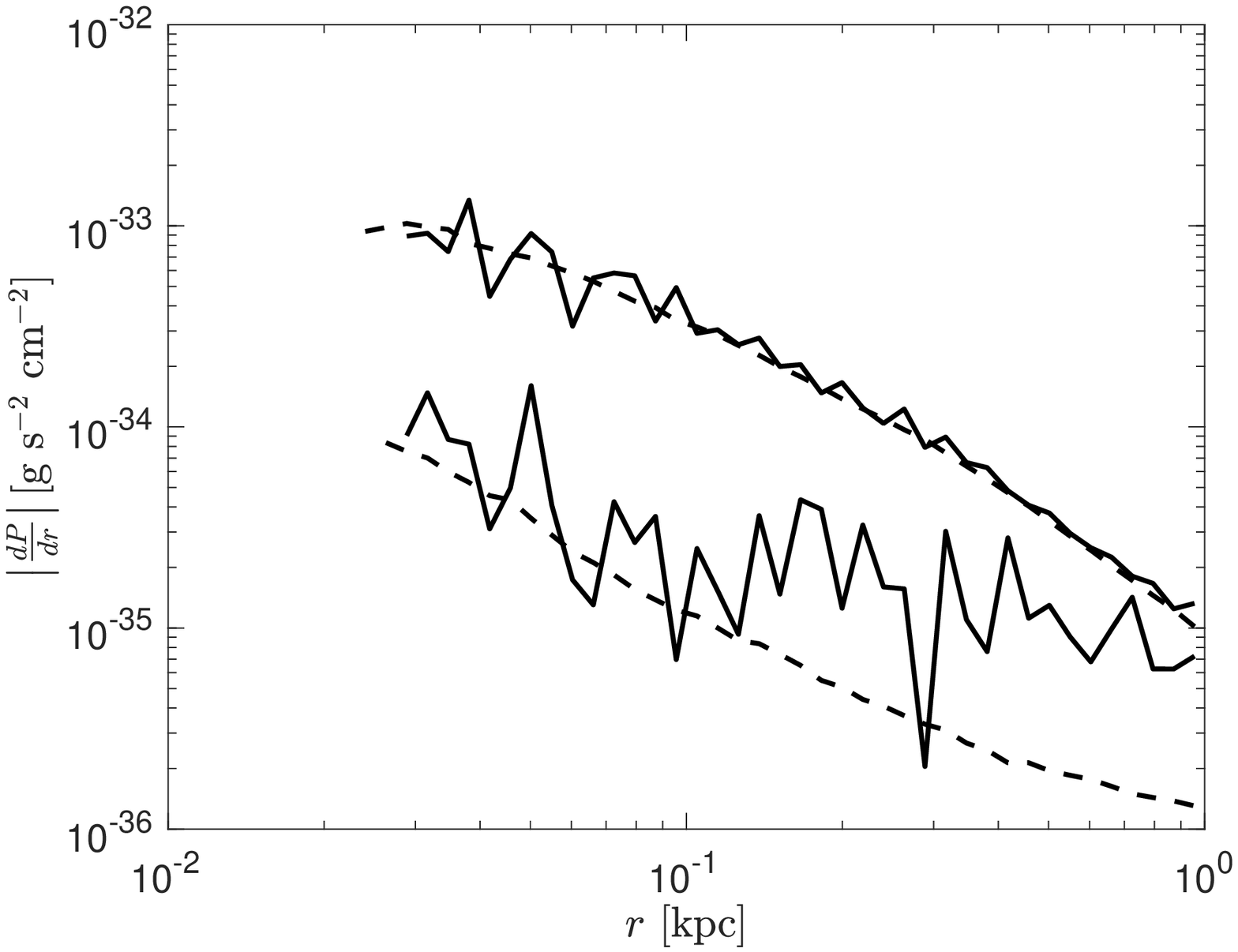}
\caption{\label{prof1_fig} The spherically averaged gas density, temperature, and HI fraction at $z=15$ (solid curves) and $z=12$ (dashed curves) for halo B subjected to ionizing flux of $F_0$ turned on at $z=20$. In the bottom right panel we plot the right-hand side (solid curves) and the left-hand side (dashed curves) of Eq. \ref{hydro_eqn} for this halo. The upper set of curves are for $z=12$ and the lower set of curves are for $z=15$. This panel demonstrates that the core is in quasi-hydrostatic equilibrium.}
\end{figure*}

At intermediate and high fluxes ($\gtrsim F_0$ for halos A and B, and $\gtrsim 0.1 F_0$ for halo C), ionization significantly delays collapse, increasing the collapse mass, by an order of magnitude or more. In these cases, the gas is ionized and photoheated to $\approx 2 \times 10^4~K$ before the formation of the halo. This photoheated gas then settles in the halo forming a pressure-supported core in quasi-hydrostatic equilibrium. This can be seen in Figure \ref{prof1_fig}, where we plot the density, ionization, and temperature profiles for halo B at $z=15$ and $z=12$ in the $F_0$ ionizing flux case. This is significantly before the collapse at $z\approx8$. In Figure \ref{prof1_fig}, we also use the spherically averaged profiles to plot each side of the equation for hydrostatic equilibrium,
\begin{equation} 
\label{hydro_eqn}
\left |\frac{dP}{dr} \right | =  \frac{G M_{\rm tot}(r) \rho_{\rm gas}(r)}{r^2},
\end{equation}
where $M_{\rm tot}(r)$ is the total mass enclosed within radius $r$, and $\rho_{\rm gas}$ is the gas density.
It is clear that the gas core in the halo is in quasi-hydrostatic equilibrium.

\begin{figure}
\includegraphics[width=85mm]{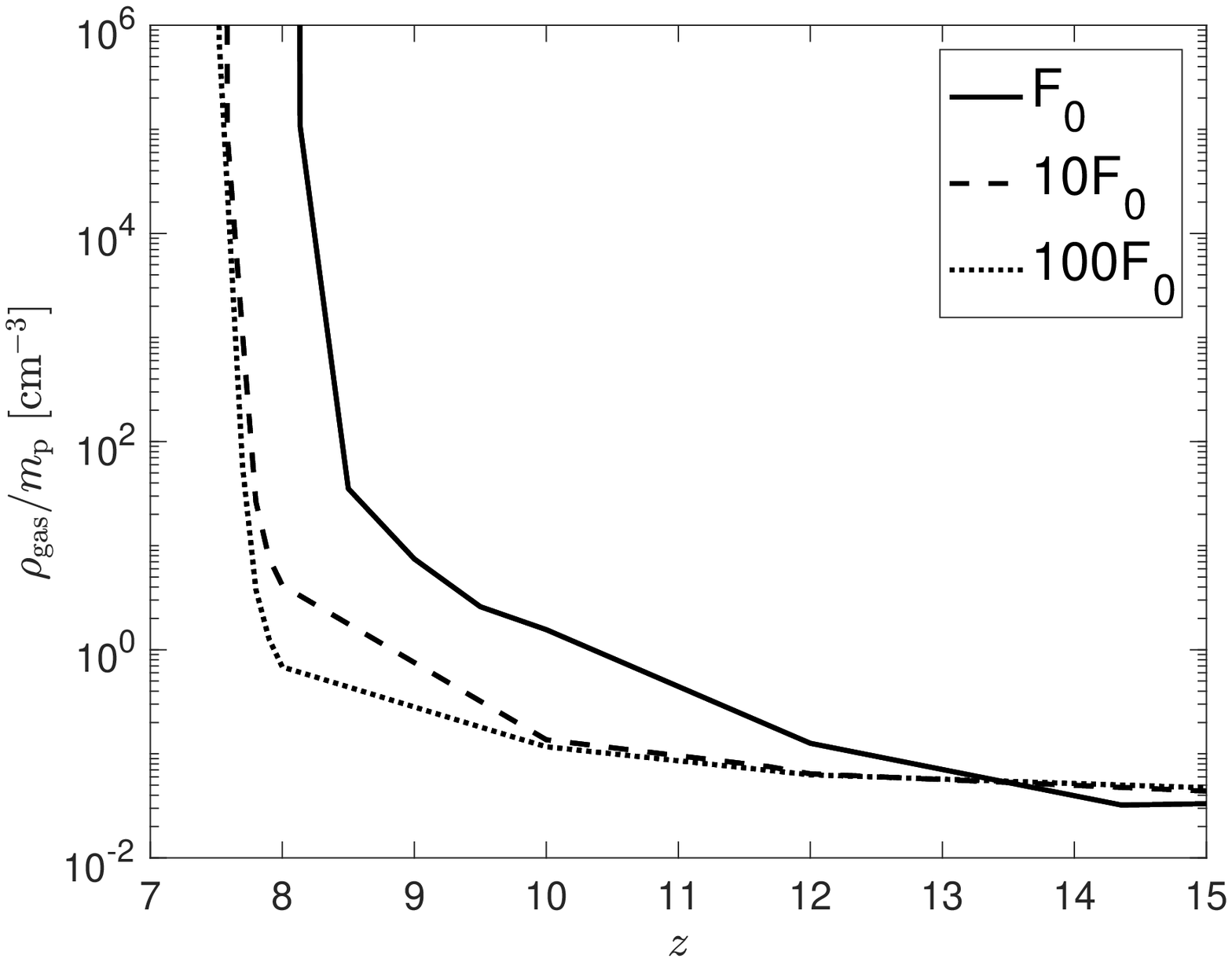}
\caption{\label{rho_vs_z_fig} The maximum gas density versus redshift for halo B. The three curves are for fluxes of $F_0$,  $10F_0$, and $100F_0$, with higher fluxes resulting in collapse at increasingly lower redshift.}
\end{figure}

As the halo increases in mass, the density of the gas core increases to maintain quasi-hydrostatic equilibrium. This can be seen in Figure \ref{rho_vs_z_fig}, where we plot the maximum density of the core in halo B as a function of redshift for a range of fluxes. As the density increases, so does the recombination rate and the neutral fraction. Eventually, the density reaches a level where the core is no longer stable and runaway collapse occurs. 

\begin{figure}
\includegraphics[width=85mm]{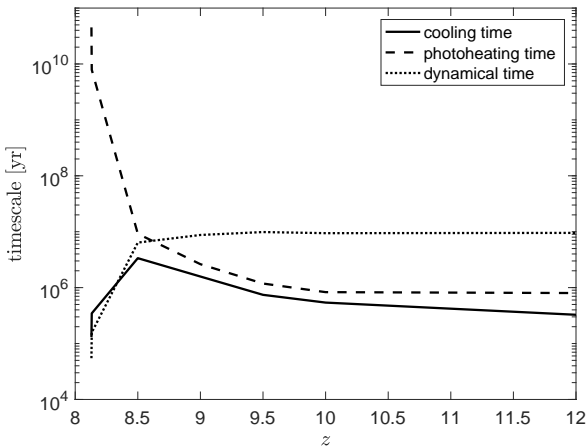}
\caption{\label{timescales_fig} { The hydrogen cooling, photoheating, and dynamical timescales as a function of redshift for halo B with $F_{\rm bg} = F_0$. These quantities are computed in the cell with the highest gas density (initially in the quasi-hydrostaticly supported gas core). Collapse is triggered when $t_{\rm p-heat} > t_{\rm dyn}$. The increase in $t_{\rm p-heat}$ leading up to collapse is driven primarily by the increased self-shielding of the gas core.}}
\end{figure}

To better understand how this collapse is triggered, we plot the photoheating, hydrogen cooling, and dynamical timescales in the core as a function of redshift in Figure \ref{timescales_fig}. We have shown the case of halo B with a background flux of $F_0$. The cooling/heating times, $t_{\rm cool}$ and $t_{\rm p-heat}$, are defined as the total thermal energy density divided by the cooling/heating rates. The dynamical time is given by the free-fall time for a spherically symmetric mass distribution, $t_{\rm dyn} =  \sqrt{\frac{3\pi}{32 G\rho_{\rm tot}}}$, where the density includes both dark matter and gas. 

At early times, we see that $t_{\rm cool} \approx t_{\rm p-heat} < t_{\rm dyn}$.  This is consistent with our earlier finding that the core is in quasi-static equilibrium and means that the core properties only evolve slowly as the halo grows in mass.  As time goes on and the potential well of the halo deepens, the dynamical time drops and the thermal equilibrium timescale increases. Once the photoheating and cooling timescales are comparable to the dynamical timescale, the core becomes unstable and collapse begins. During collapse, the cooling time is roughly equal to the dynamical time and compressional heating is balanced by hydrogen cooling. Thus, $t_{\rm p-heat} \gtrsim t_{\rm dyn}$ is the key criterion for collapse to be triggered. 

Suppressing numerical factors and physical constants, the heating timescale goes as 
\begin{equation}
t_{\rm p-heat} \propto \frac{T}{f_{\rm sh}  F_{\rm bg} \mu f_{\rm HI} },
\label{theat_eqn}
\end{equation}
 where $\mu$ is the mean molecular weight, which depends on the neutral fraction, $f_{\rm HI}$. Leading up to collapse for the case plotted in Figure \ref{timescales_fig}, the temperature, dark matter density, and neutral fraction (which is $f_{\rm HI} \approx 1)$ in the core do not vary quickly. Thus, the increase in $t_{\rm p-heat}$ is driven by the increase in self-shielding, which is changed mostly by the increasing gas density. We note that in the case where $f_{\rm HI} \ll 1$, replacing the equilibrium value for $f_{\rm HI}$ gives us a heating time scaling of $t_{\rm p-heat} \propto T^{3/2}/\rho_{\rm gas}$. Thus, the reason the increased self-shielding raises $t_{\rm p-heat}$ in the present example is due to the fact that the gas becomes mostly neutral well before collapse.

Generally, the density, ionization fraction, and temperature evolution of the gas core determines the point at which the collapse criterion ($t_{\rm p-heat} \gtrsim t_{\rm dyn}$) is satisfied. In principle, one may be able to analytically estimate the equilibrium values of these quantities at each redshift by solving the equations of hydrostatic equilibrium in a fixed dark matter potential. However, the precise results may depend on the chosen boundary conditions. We defer these calculations to future work. 

From Eq. \ref{theat_eqn}, it is clear that self-shielding has an important impact on when collapse occurs. This has been verified in our simulations. We repeated the simulation of halo B with $F_{\rm bg} = F_0$, but without any self-shielding (i.e. $f_{\rm sh} = 1$) and found that collapse was delayed. In this case, collapse occurs approximately at $z=7.3$, which is even slightly later than in the case of $F_{\rm bg} = 100F_0$ when self-shielding is included.

We note that our simulations do not include pressure from Ly$\alpha$ radiation produced during cooling. In principle, this could provide some additional support against gravity, delaying collapse to larger masses. In \cite{2010ApJ...712L..69S}, one-zone models were utilized to analyze the collapse of pristine gas including Ly$\alpha$ trapping (see also \cite{2006ApJ...652..902S}). They find that for increased optical depth to Ly$\alpha$ scattering, the  overall hydrogen cooling rate is not greatly changed (see their Figure 3). This is because at high density two-photon decays lead to cooling through optically thin continuum photons. That there is a not a significant reduction in the overall cooling rate suggests that hydrogen cooling radiation can escape in a time scale similar to the dynamical time or faster. Thus, due to this relatively short trapping time scale, we do not expect Ly$\alpha$ pressure to have a large impact on the results presented above. We defer more detailed study of this effect to future work.

\begin{figure*}
\includegraphics[width=80mm]{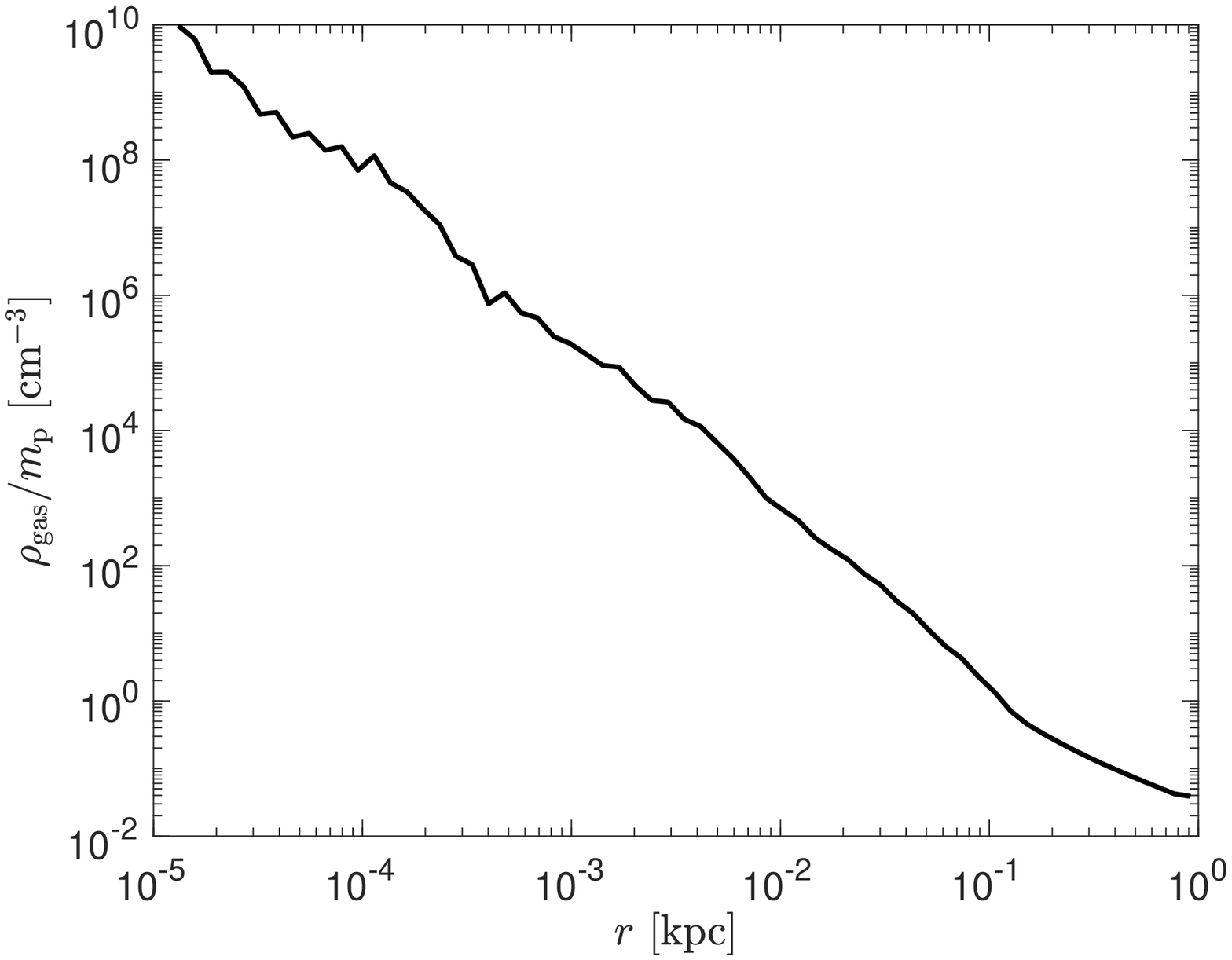}
\includegraphics[width=80mm]{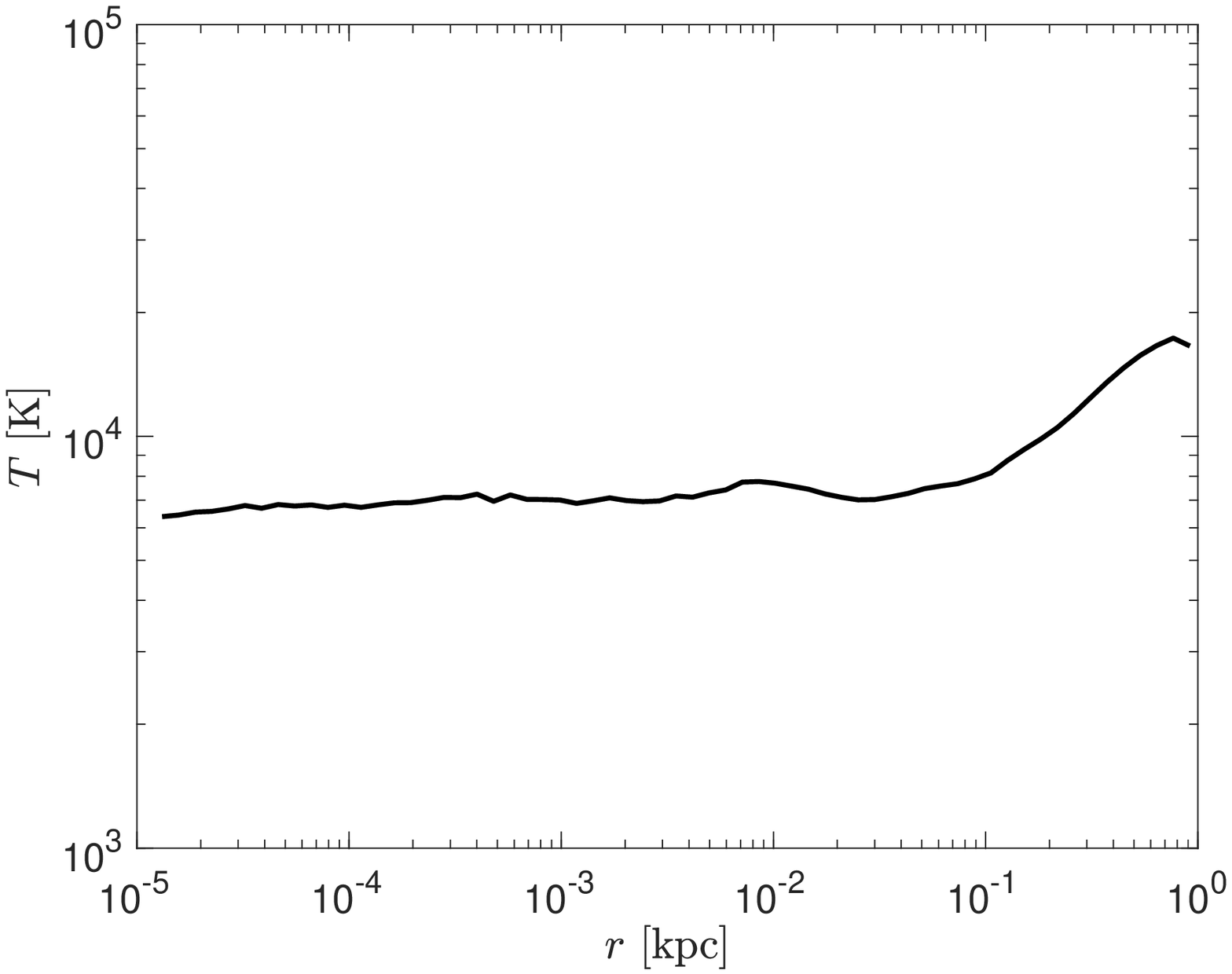}
\includegraphics[width=80mm]{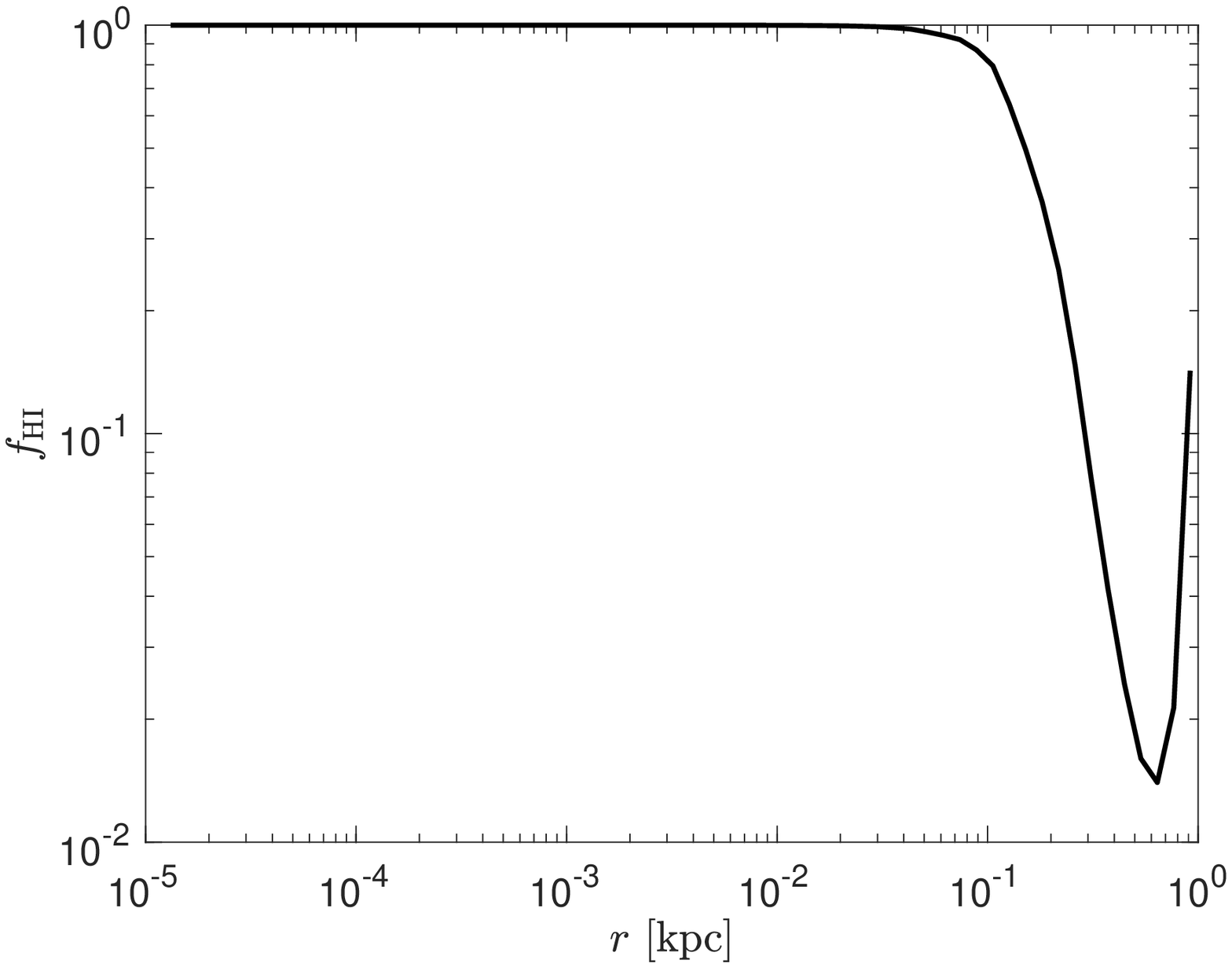}
\includegraphics[width=80mm]{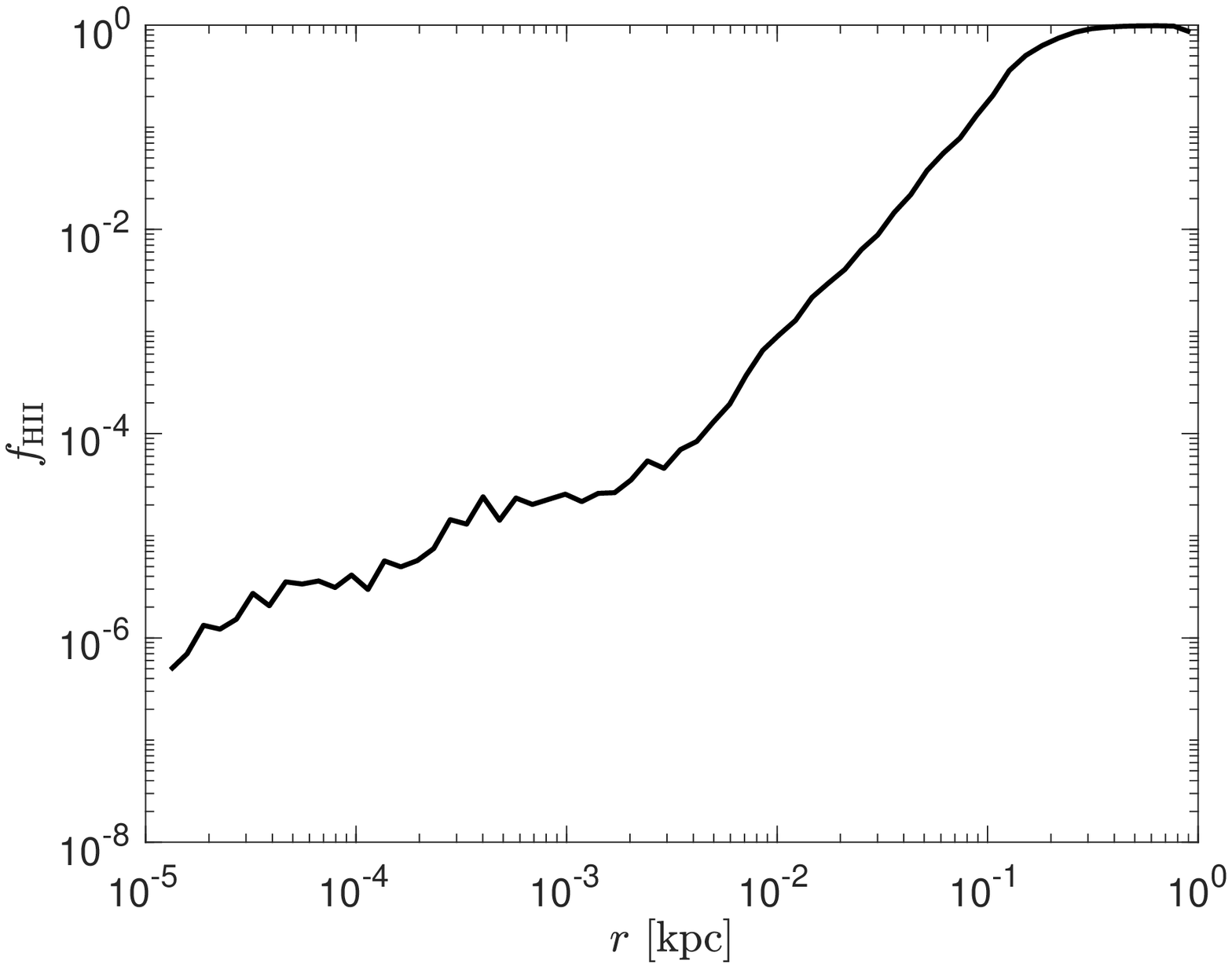}
\caption{\label{prof2_fig} The spherically averaged density, temperature, HI fraction, and HII fraction of halo B subjected to an ionizing flux of $F_0$ at the redshift of runaway collapse ($z= 8.1$).}
\end{figure*}

For the highest fluxes we simulate, the collapse mass depends very weakly on flux. This is because once a halo reaches a very high mass, corresponding roughly to the virial temperature surpassing the photoheated gas temperature, the density of the core rapidly increases. The density quickly grows so large that even for our most extreme flux, the gas rapidly recombines and collapses.  In Figure \ref{prof2_fig}, we plot an example of the density, ionization fraction, and temperature profiles at collapse.

\section{Limits on Pop III Stellar Mass}
\begin{figure*}
\includegraphics[width=80mm]{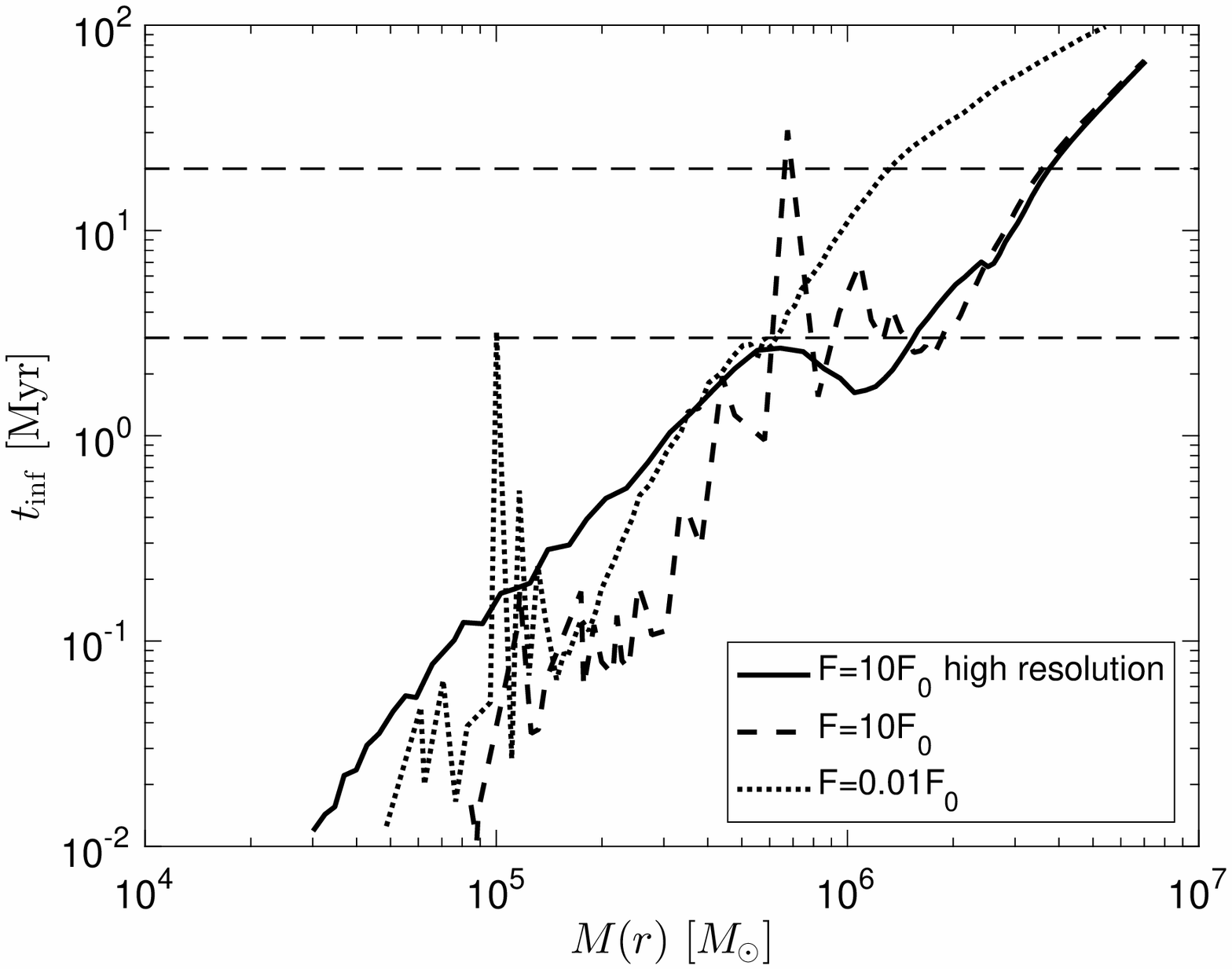}
\includegraphics[width=80mm]{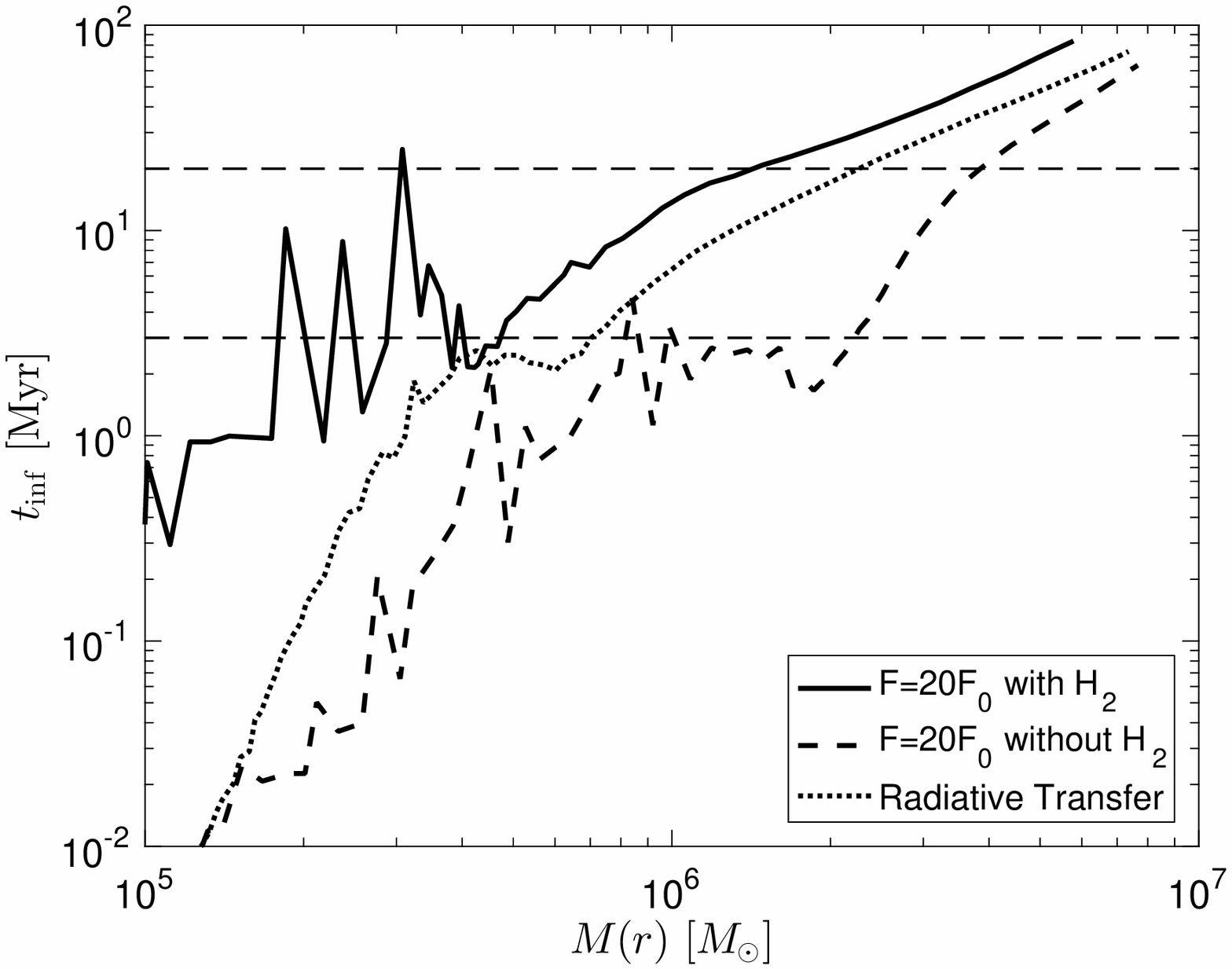}
\caption{\label{stellar_fig1}
  The infall time, inferred from the spherically averaged radial velocity,
  versus enclosed gas mass for halo A. In the left panel we show the impact of resolution and in the right panel we show the effects of our local self-shielding approximation (all cases use this approximation except for the one labelled ``Radiative Transfer'') and including molecular hydrogen.    }
\end{figure*}

\begin{figure*}
\includegraphics[width=88mm]{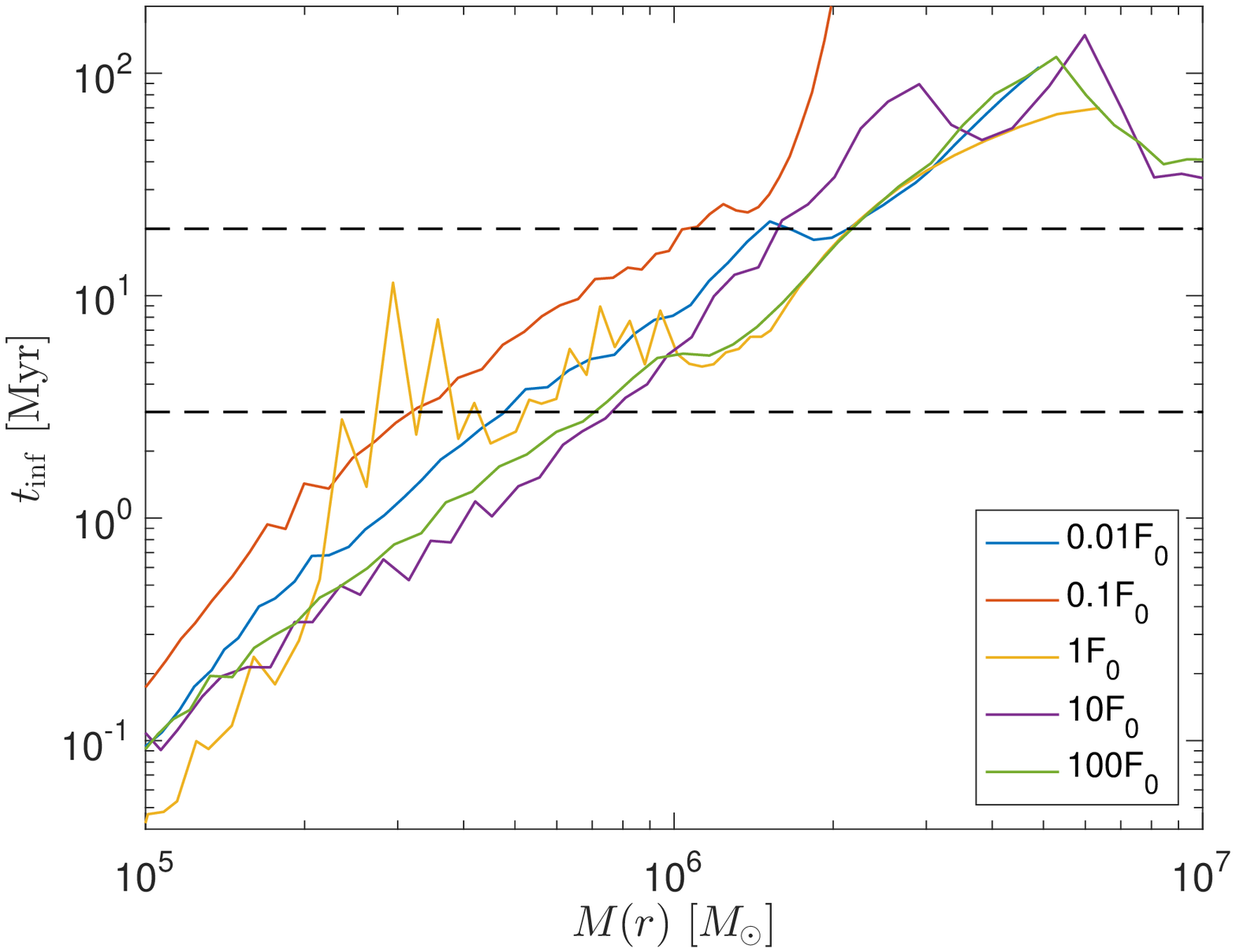}
\includegraphics[width=88mm]{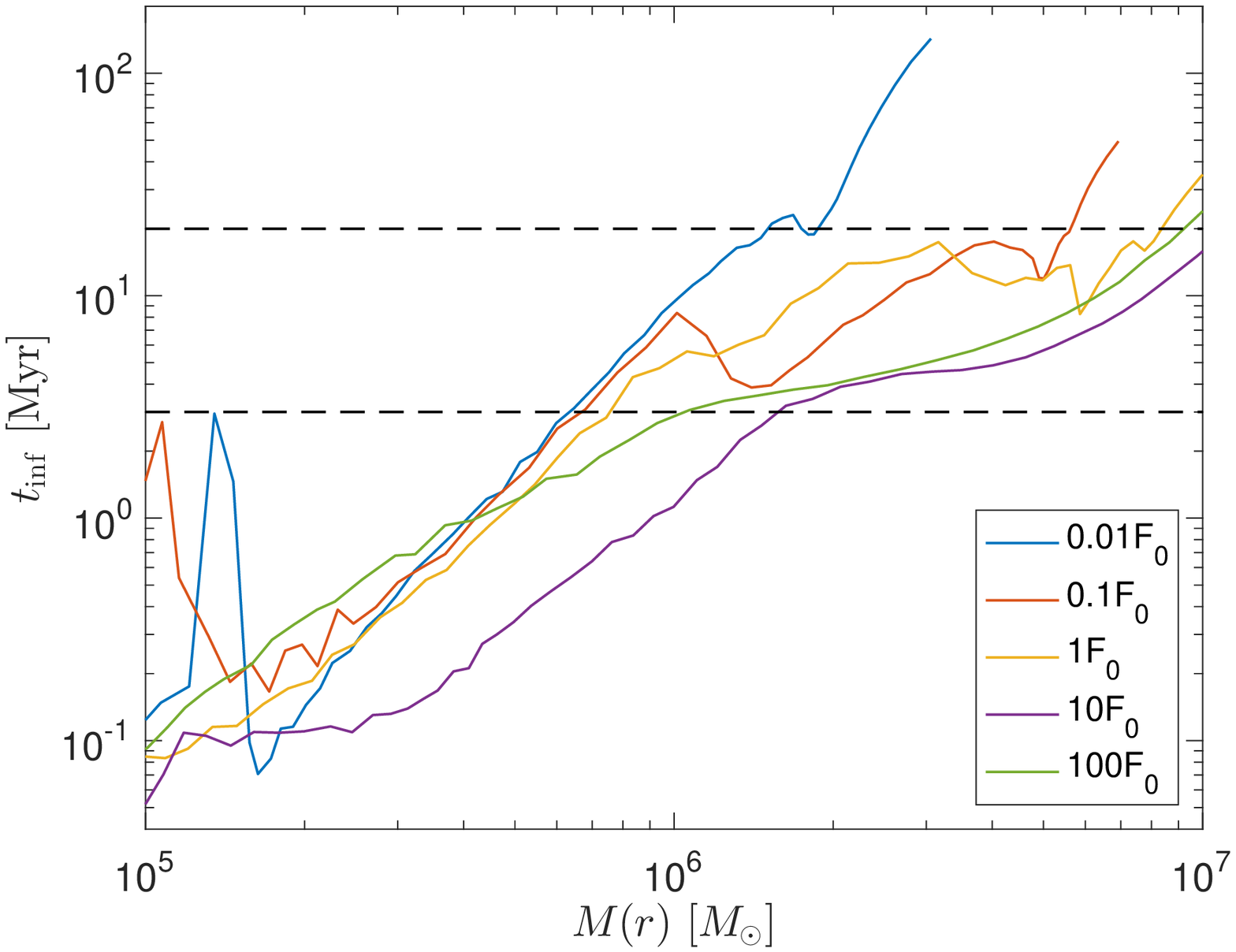}
\caption{\label{stellar_fig2}
  The infall time versus enclosed gas mass for halos B (left panel) and C (right panel) for a range of ionizing fluxes.  }
\end{figure*}

In this section, we discuss the total mass of a Pop III starburst which can form in dark matter halos subjected to photoionization feedback. While simulating the detailed processes of star formation is beyond the scope of this paper, we estimate upper limits on the Pop III stellar mass by examining the radial velocity profiles of the gas at collapse. This is shown in Figures \ref{stellar_fig1} and \ref{stellar_fig2}, where we plot the infall time $t_{\rm inf} = r/v_{\rm rad}$ (where $v_{\rm rad}$ is the spherically averaged radial velocity), versus the gas mass enclosed within radius $r$. This shows how much gas can reach the central region of a halo for a range of timescales. We focus mostly on how much gas can reach the center by $3~{\rm Myr}$, because this is the lifetime of $\sim 80~M_\odot$ Pop III stars \citep{2002A&A...382...28S}. After this, gas may be polluted by  metals from the first stellar generation, and subsequently form Pop II stars. For completeness, we also consider $20~{\rm Myr}$, which corresponds to the lifetime of a $9~M_\odot$ Pop III star \citep{2002A&A...382...28S}. We note that this timescale is probably too long to be relevant because not having stars with masses $\gtrsim 100~M_\odot$ would require an unrealistically bottom-heavy initial mass function (IMF). Keeping the star-forming gas pristine for this long would require very inefficient metal mixing.

In Figure \ref{stellar_fig1}, we plot infall times for halo A. In the left panel we show the impact of resolution. For strong ionizing feedback, our high- and normal-resolution runs agree well above $M(r)\approx 10^6 M_\odot$ (corresponding to $t_{\rm inf} \approx 3 ~{\rm Myr}$). Below this value, our normal resolution contains more noise and shows a factor of a few larger mass at a given infall time. For comparison we also include a case with weak feedback and find that a factor $\sim 3$ less gas can reach the center within a few Myr. 

In the right panel of Figure \ref{stellar_fig1}, we show the impact of including/excluding molecular hydrogen (with a $J_{LW}=100\times J_{21}$ LW background) and of our local self-shielding approximation. We find that the self-shielding approximation and the full radiative transfer simulation with molecular hydrogen give very similar results. However, compared to not including molecules, molecular cooling decreases the mass which can reach the center significantly (e.g. factor of $\sim 4$ for $3~{\rm Myr}$, from $\approx 2\times 10^6 M_\odot$ to $\approx 5\times 10^5 M_\odot$ ). Thus, for the results shown for halos B and C discussed next, the upper limits are rather conservative and could be a factor of a few lower than what we find without including molecular hydrogen. We suspect that molecular hydrogen increases the infall time because it cools the gas to lower temperatures, reducing the sound speed and creating weak shocks at lower velocities which slows this gas. This is consistent with the results of \cite{2010MNRAS.402.1249S}, which show that the infall velocity in a collapsing atomic cooling halo is approximately equal to the sound speed.

In Figure \ref{stellar_fig2}, we plot the infall time for halos B and C over a large range in ionizing flux. We find that within $3~ {\rm Myr}$, our limiting gas mass is between $\approx 2\times 10^5 - 2\times 10^6~ M_\odot$. For this infall time, we find that strong ionizing radiation can increase the infalling mass by a factor of a few, but with scatter from run to run. Interestingly, if we allow additional time for star formation (due to longer-lived stars or slower metal mixing), we can reach significantly higher masses. For example, when halo C is exposed to flux $\gtrsim F_0$, approximately $10^7~M_\odot$ of gas can reach the center of the halo within $\sim 20~{\rm Myr}$, approximately 5 times more than in the lowest flux shown.

Finally, we consider the observational prospects of Pop III stars formed in massive halos due to LW and photoionization feedback. Of the three halos we simulate, halo C seems the most likely to produce an observable starburst due to its higher infall mass than halo B and lower redshift than halo A. For strong feedback, $\sim 10^6~{\rm M_\odot}$ and $\sim 10^7~{\rm M_\odot}$ of gas can fall to the center of halo C within 3 Myr and 20 Myr, respectively. We note that including molecular hydrogen would likely reduce these masses by a factor of a few. 
\cite{2011ApJ...740...13Z} computed the total mass of a Pop III galaxy which could be observed in a 100-hour integration with \emph{JWST}. They found that within 3 Myr of a starburst $\sim {\rm a~few~} \times 10^4~ M_\odot$ and  $\sim {\rm a~few~} \times 10^5~ M_\odot$ of Pop III stars could be detected at $10-\sigma$ for maximal and no-nebular flux contribution, respectively (at $z \sim 6$). Thus, if a significant fraction of the gas which makes it to the center of halo C forms Pop III stars, it could potentially be observable with \emph{JWST}. 

Halo C subjected to an ionizing flux of $0.1F_0$ is similar to the case of photoionization feedback by a $6.6 \times 10^{11}~M_\odot$ star-forming dark matter halo discussed in \cite{2016MNRAS.460L..59V}. It was estimated that a comoving number density of $\approx 10^{-7}~{\rm Mpc}^{-3}$ of Pop III starbursts would be visible at $z=6.6$ if star formation was suppressed by photoionization feedback up to $\sim 10^9 ~M_\odot$ in halo mass. Since we find that star formation occurs in smaller halos, we expect a higher space density.  We note that a number density of $10^{-7}~{\rm Mpc}^{-3}$ corresponds to $\approx 0.002$ per \emph{JWST} field of view per unit redshift at $\approx6$. Thus, it may not be possible to find Pop III galaxies in halos much more massive than the atomic cooling threshold in blind searches. A better strategy is likely to be searching near massive galaxies which have already been discovered. Pristine halos with $T_{\rm vir}\approx 10^4~{\rm K}$ are likely to be more common (since they only require strong LW radiation),
and may form similar masses of Pop III stars (perhaps a factor of a few less as suggested by the limits discussed above).

Our limits suggest that CR7 cannot be explained by Pop III stars alone, as its luminosity requires $\sim 10^7 ~ M_\odot$ of massive Pop III stars. However, photoionization feedback could have played a role in keeping its metallicity low for much of its halo formation history.

\section{Discussion and Conclusions}
We have performed cosmological zoom-in simulations to study the delay of Pop III star formation due to strong photoionization feedback.  We utilize a local self-shielding approximation based on \cite{2013MNRAS.430.2427R}, which for our test case is consistent with the ray tracing radiative transfer available in \textsc{enzo}, but is much less computationally demanding. Using this approximation, we determine the redshift and halo mass of runaway gas collapse immediately preceding star formation for three different dark matter halos over a wide range in ionizing fluxes. 

We find that for low ionizing flux, collapse first occurs at $T_{\rm vir} \approx 10^4~{\rm K}$. This is because it is the lowest temperature where atomic hydrogen cooling is efficient (we assume molecular hydrogen is photodissociated by a strong LW background). At intermediate flux, the collapse is delayed significantly, with the length of the delay depending on the strength of the ionizing flux. For halos A and B, feedback becomes important at  $F_0$, while for halo C it becomes important at $0.1F_0$. These high fluxes (required to exist already at high redshift, far in advance of collapse) suggest that for ionization feedback to become important, a halo is required to be near one or more large star forming galaxies. At fluxes $\gtrsim 10F_0$, the  threshold halo mass at collapse reaches a plateux and depends only weakly on flux, with stars forming in halos
of a  $\approx 5\times 10^8 ~ M_\odot$.
This is roughly an order of magnitude higher than the low-flux case (corresponding to a factor of 5 higher in $T_{\rm vir}$). We note that while we have only considered redshift-independent fluxes (with some turn-on time), it may be possible for the flux to be lower at higher redshifts, ramping up over time, and still result in similar delays of star formation. We leave an investigation of different flux histories for future work.

We note that for an escape fraction of 0.1 in ionizing photons and 1 in LW photons, the LW background corresponding to $F_0$ is $J_{\rm LW} = 150 \times J_{21}$. For flux values $\gtrsim 10F_0$, the LW background could exceed the critical intensity required to form a supermassive star, leading to a so-called direct collapse black hole (DCBH) \citep{2003ApJ...596...34B,2016arXiv160902142W,2010MNRAS.402.1249S,2014MNRAS.445..544S}. Thus, for these extremely high fluxes it is unclear what the final state of the collapsing gas will be, since DCBH formation has not been studied in this context of massive halos and strong and long-lived ionizing radiation. \cite{2016MNRAS.459.3377R} studied DCBH formation including photoionization, but in smaller halos and with a shorter duration of ionizing radiation than we consider here. 

Previously it has been suggested that the mass at which star formation is suppressed is set by the Jeans mass of photoheated gas at the cosmic background density, because below this mass pressure forces support the gas against the force of gravity. We find that star formation can occur at much smaller masses. For reference, the Jeans mass of $T=2\times 10^4 ~{\rm K}$ gas at the mean cosmic density at $z=6$ is $M_{\rm J}  \sim 3 \times 10^{10} M_\odot$. Assuming the gas was locally ionized in the distant past, the filtering mass \citep{2000ApJ...542..535G} is $M_{\rm F}  \sim 5 \times 10^{9} M_\odot$, including the factor of 8 correction from \cite{2009MNRAS.399..369N} (see Eqs. 2 and 8 in \citealt{2014MNRAS.444..503N}). Even for extremely high ionizing fluxes, we find about an order of magnitude lower halo masses at collapse. 

By examining the outputs of our simulations, we were able to understand the physical processes which set the collapse mass. For strong ionizing feedback, after the gas is ionized, it settles into an ionized core in quasi-hydrostatic equilibrium. As the halo grows, the density of the core increases leading to an increasing recombination rate.
Eventually, the photoheating timescale in the self-shielded core becomes shorter than the dynamical timescale, leading to runaway collapse and star formation.
For intermediate fluxes, the higher the flux, the longer this takes to happen. Once the virial temperature of the halo significantly exceeds the temperature of the photoheated gas, the density increases rapidly and even very large flux cannot delay collapse beyond $T_{\rm vir}\approx 3\times 10^4 ~{\rm K}$.

To study the total stellar mass of Pop III starbursts, we examine the radial velocity profiles of collapsing gas at the end of our simulations. This allows us to put upper limits on the star formation which can occur within a given duration of time after collapse. Our halo C represents the most promising possibility of an observable mass of Pop III stars. We find that for strong feedback, $\sim 10^6~{\rm M_\odot}$ and $\sim 10^7~{\rm M_\odot}$ of gas reach the center of the halo C within 3 Myr and 20 Myr (corresponding to the lifetimes of 80 and 9 $M_\odot$ stars, respectively). Including molecular hydrogen would likely reduce these numbers by a factor of a few. On the other hand, photo-heating by sources with a harder spectrum to higher temperatures could potentially increase the masses. However for the spectra considered here, we point out that photoionization flux increases the halo mass of collapse much more than the possible star forming gas. 

These Pop III galaxy mass limits suggest that CR7 cannot be explained solely by a Pop III starburst formed from photoionization feedback (with a Pop II ionizing source), as this would require $\approx 10^7 ~M_\odot$ of massive ($\gtrsim 100 ~M_\odot$) Pop III stars to generate the HeII 1640 \AA\ emission line \citep{2015ApJ...808..139S, 2015MNRAS.453.2465P, 2016MNRAS.460L..59V}. This is consistent with the recent observations by \cite{2016arXiv160900727B}, who find evidence of oxygen lines in Spitzer/IRAC data. 

If a significant fraction of the infalling gas mass we find in our simulations forms Pop III stars, they may be observable with future telescopes such as \emph{JWST}. Our simulations suggest that there should not be a dramatic difference in total stellar mass between halos with strong or weak ionizing feedback (assuming both have molecular cooling suppressed by LW radiation). Pop III starbursts formed near bright galaxies are likely to be a factor of a few times more massive than those with weak ionizing feedback, but may be hosted by over an order of magnitude more massive dark matter halos. Thus, the ratio of stellar mass to halo mass is significantly lower in halos which experienced strong ionizing feedback. In principle, this may be observable (through e.g. differences in the velocity dispersion of stars).

In future work, it will be important to simulate the detailed process of star formation
and internal metal mixing after the initial collapse to better estimate the total number and mass of Pop III stars formed. It will also be useful to perform a detailed estimate of the cosmic number density of these Pop III starbursts which will be observable with \emph{JWST} or future ground-based telescopes such as the \emph{Thirty Meter Telescope} and the \emph{European Extremely Large Telescope}.

\section*{Acknowledgements}
We thank John Regan for useful discussions. 
The Flatiron Institute is supported by the Simons Foundation. EV was
also supported by the Columbia Prize Postdoctoral Fellowship in the
Natural Sciences.  ZH was supported by NASA grant NNX15AB19G and by a
Simons Fellowship in Theoretical Physics.  GLB was supported by NASA
grant NNX15AB20G and NSF grants AST-1312888 and AST-1615955.  The
computations in this paper were carried out on the NASA Pleiades
supercomputer.

\bibliography{paper.bib}

\end{document}